\ifdefined\ONECOLUMN
\documentclass[journal, 12pt, onecolumn, draftclsnofoot, letterpaper]{IEEEtran}
\else
\documentclass[conference, a4paper]{IEEEtran}
\fi

\IEEEoverridecommandlockouts

\usepackage{amsmath,amssymb,mathrsfs,amsbsy}
\usepackage{cite}
\usepackage{graphicx,color}
\usepackage[bookmarks=false]{hyperref}
\usepackage{psfrag}
\usepackage{subfigure}
\usepackage{url}
\usepackage{balance}
\usepackage{stfloats}

\usepackage{array}
\usepackage{fancyhdr}
\usepackage{float}
\usepackage{epsfig}
\usepackage[nomain,acronym,toc]{glossaries}
\usepackage{algorithm,algorithmic}
\usepackage{caption}
\usepackage{tikz,siunitx}
\usepackage{makecell}

\newtheorem{thm}{Theorem}

\newtheorem{lem}{Lemma}

\renewcommand{\eqref}[1]{(\ref{#1})}
\definecolor{sblue}{RGB}{0,51,160}
\definecolor{seaBlue}{RGB}{0,105,148}

\ifCLASSINFOpdf
\else
\fi

\makeatother

\begin{document}
\title{\huge	
On Chernoff Lower-Bound of Outage Threshold for Non-Central $\chi^2$-Distributed MIMO Beamforming Gain
}
\author{
Jinfei Wang, Yi Ma, and Rahim Tafazolli\\
{\small 5GIC and 6GIC, Institute for Communication Systems, University of Surrey, Guildford, UK, GU2 7XH}\\
{\small E-mails: (jinfei.wang, y.ma, r.tafazolli)@surrey.ac.uk}\\
}


\maketitle

\begin{abstract}	
The cumulative distribution function (CDF) of a non-central $\chi^2$-distributed random variable (RV) is often used when measuring the outage probability of communication systems.
For adaptive transmitters, it is important but mathematically challenging to determine the outage threshold for an extreme target outage probability (e.g., $10^{-5}$ or less). 
This motivates us to investigate lower bounds of the outage threshold, and it is found that the one derived from the Chernoff inequality (named Cher-LB) is the most {effective} lower bound. 
The Cher-LB is then employed to predict the multi-antenna transmitter beamforming-gain in ultra-reliable and low-latency communication, concerning the first-order Markov time-varying channel.  
It is exhibited that, with the proposed Cher-LB, pessimistic prediction of the beamforming gain is made sufficiently accurate for guaranteed reliability as well as the transmit-energy efficiency.
\end{abstract}


\IEEEpeerreviewmaketitle

\section{Introduction}\label{secI}

Consider $K$ independent real-Gaussian random variables (RVs), $(\alpha_0, \alpha_1, ..., \alpha_{K-1})$, with $\alpha_k\sim\mathcal{N}(\mu_k,\sigma^2)$, $_{k=0,...,K-1}$, where $\mu_k$ and $\sigma^2$ stand for their means and variances, respectively.
Define a RV, $\beta$, as follows
\begin{equation}\label{eqn01}
\beta\triangleq\sum_{k=0}^{K-1}\alpha_k^2.
\end{equation}
It is understood that $\beta$ obeys the non-central $\chi^2$-distribution with the following cumulative distribution function (CDF) \cite{AP91}
\begin{IEEEeqnarray}{rl}\label{eq02}
F_\beta(x)=1-\mathcal{Q}_{\mathrm{M},\frac{K}{2}}(\mathcal{M}/\sigma,\sqrt{x}/\sigma),
\end{IEEEeqnarray}
where $\mathcal{Q}_{\mathrm{M},\frac{K}{2}}$ stands for the generalized Marcum Q-function of order $(K)/(2)$ \cite{Andras2010}, and $\mathcal{M}\triangleq\sqrt{\sum_{k=0}^{K-1}\mu_k^2}$.

We are interested in the probability when $\beta$ is smaller than a threshold $\beta_{\mathrm{T}}$, denoted by $\mathscr{P}(\beta<\beta_{\mathrm{T}})$. 
When this probability is required to be a specific value, i.e., $\epsilon\sim(0,1)$, can we find the closed-form of $\beta_{\mathrm{T}}$? 
If the answer is `{\em No}', can we find a good (or tight) lower-bound of $\beta_{\mathrm{T}}$ (denoted by $\beta^\perp\leq\beta_{\mathrm{T}})$?
This problem is important in the scope of signal processing for communications with many applications such as in spectrum sensing, diversity techniques, precoding, etc..  
More remarkably, in multiantenna (i.e., MIMO) empowered ultra-reliable low-latency communication (URLLC) systems, 
the transmitter must make sufficiently accurate prediction of MIMO beamforming gain to ensure extremely-low outage probability even for a single transmission \cite{8472907}.
This technical requirement has been confirmed by telecommunication stakeholders such as Nokia Bell Labs and so forth \cite{8723572,9114878}. 

It is worth mentioning that the reverse problem of our study, i.e., evaluating $\epsilon$ based on $\beta_{\mathrm{T}}$, is a classical problem in communication.
Although various bounds of the Marcum Q-function have been utilized to simplify the reverse problem (e.g., \cite{Simon2000}), they are not directly applicable to our study, since the evaluation of $\beta_\mathrm{T}$ based on $\epsilon$ remains transcendental with these bounds. 
This distinction separates our study from its reverse problem.

\subsection{Prior Art}
Mathematically, the problem requires to solve the following equation
\begin{equation}\label{eqn04}
\epsilon=F_\beta(\beta_{\mathrm{T}})\int_{0}^{\beta_{\mathrm{T}}}f_\beta(x)dx,
\end{equation} 
where $f_\beta(x)$ stands for the probability density function (PDF) of $\beta$.
To the best of our knowledge, there is so far no closed-form solution available in the literature. 
Moreover, we are only interested in low-complexity approaches to evaluating $\beta_{\mathrm{T}}$.
This is because the evaluation of $\beta_{\mathrm{T}}$ becomes latency-sensitive for real-time signal processing when the parameters (e.g., $\mathcal{M}$) are time-varying in communication scenarios.  

Setting aside that $\epsilon$ is extremely small in URLLC systems, a couple of solutions to \eqref{eqn04} have been presented in mathematical studies. 
These include Abdel-Aty's first/closer approximations \cite{AbdelAty1954}, and Sankaran's $z1$/$z2$ approximations \cite{Sankaran1963}.
The idea is to approximate a scaled version of $\sqrt{\beta}$ (or $\sqrt[3]{\beta}$) as a Gaussian RV.
Then, the scaled $\sqrt{\beta_\mathrm{T}}$ is easily obtained through the inverse Gaussian Q-function.
However, the accuracy of these approximations could be questionable when $\epsilon$ is extremely small (i.e., $\beta_{\mathrm{T}}$ is small).
This is because when $\beta_{\mathrm{T}}$ is small, it could be wrongly deemed to be negative when approximated as Gaussian.

In light of this, recent communication studies have turned to finding the lower-bound of $\beta_{\mathrm{T}}$ (denoted by $\beta^\perp$).
The most widely studied case is when $\beta$ is central $\chi^2$-distributed (i.e., $\mathcal{M}^2=0$) (e.g., \cite{8660712,9120745}), where the maximal ratio combining (MRC) is adopted to enhance the receiver-beamforming gain in Rayleigh fading channel. 
In this case, the polynomial lower-bound (Poly-LB) in \cite{8660712} is notable for its compact mathematical form and good tightness.
The idea of the Poly-LB is to expand polynomial series of $F_\beta(\beta_{\mathrm{T}})$ and retain the first term.
When $\mathcal{M}>0$, this compact form can be preserved based on \cite{Andras2010}.
Here we extend the discussion in \cite{8660712}, which is a special case of $(K=2)$, to the general case:
\begin{thm}\label{thm03}
Consider a RV, $\beta$, following the non-central $\chi^2$-distribution specified in \eqref{eq02}. 
The Poly-LB form of $\beta_{\mathrm{T}}$ (denoted by $\beta^\perp_{\text{poly}}$) is given by
\begin{equation}\label{eqn11}
\beta^\perp_\text{poly}=2\sigma^2(\epsilon\Gamma(K/2+1) )^{\frac{2}{K}}\exp\left(\frac{\mathcal{M}^2}{K\sigma^2}\right).
\end{equation}
\end{thm}
\begin{IEEEproof}
The proof is abbreviated for space limit.
\end{IEEEproof}

Nonetheless, \textit{Theorem~\ref{thm03}} implies $\beta^\perp_{\text{poly}}\to\infty$ as $\sigma^2\to0$, whereas it is expected that $\beta^\perp_{\text{poly}}\to\mathcal{M}^2$ as $\sigma^2\to0$. This indicates that although the Poly-LB preserves its compact form, it may no longer serve as a lower-bound, as it does in the central case.

There are also other commonly used mathematical tools to derive lower/upper bound of error probabilities. 
Those include the Jensen's inequality (e.g., \cite{7727938}), Chebyshev inequality (e.g., \cite{1650344}) as well as the Chernoff inequality \cite{9120745}. 
Specifically, Jensen's inequality relies on the convexity of $f_\beta(x)$, which is not always available. 
The Chebyshev inequality evaluates the probability of $|\beta-\mathbb{E}(\beta)|$, where $\mathbb{E}(\cdot)$ stands for the mean of a RV, $|\cdot|$ for the Euclidean norm.
However, this may result in significant inaccuracy as $f_\beta(x)$ is highly asymmetric.
The Chernoff inequality, on the other hand, has been used as lower-bound (i.e., the Cher-LB) in the central case \cite{9120745}.
Therefore, we focus on the comparison between the approximations \cite{AbdelAty1954,Sankaran1963}, the Poly-LB \cite{8660712}, and the Cher-LB in our results.
It is worth highlighting that the discussions in \cite{9120745} is not straightforwardly applicable to the context of non-central $\chi^2$-distribution;
please see our discussion in Sec.~\ref{secII}.

\subsection{Contribution}

{In Sec.~\ref{secII}}, we first study the Cher-LB ($\beta^\perp_{\text{Cher}}$), and compare it to the aforementioned studies in the literature.
It is found that for $\epsilon\in(0,1)$, $\beta^\perp_{\text{Cher}}$ monotonically increases with $\epsilon$ and falls into the range of $(0,\mathbb{E}(\beta))$ with $\mathbb{E}(\beta)=\mathcal{M}^2+K\sigma^2$.
Therefore, $\beta^\perp_{\text{Cher}}$ can be obtained through line-searching (see \textit{Theorem~\ref{thm05}} and \textbf{Algorithm~\ref{agthm2}}).
{Numerical results demonstrate that the Cher-LB is the most effective estimation of $\beta_\mathrm{T}$,} i.e., it is the only approach that always fulfill the outage requirement regardless the change of $\mathcal{M}$.
These findings establish the Cher-LB as a valid and valuable enabler for MIMO-URLLC.

{Then, in Sec.~\ref{secIII}, the Cher-LB is} employed to enable MIMO beamforming-gain prediction in URLLC systems, concerning the first-order Markov time-varying channel.
Due to the channel time-variation, there is only imperfect channel knowledge available at the transmitter (i.e., imperfect CSI-T). 
To prevent the system from being interference-dominated, where diversity combining is hardly useful {(see Sec.~\ref{secIIIa})}, we focus on single-stream MIMO transmission for URLLC.
In this case, the beamforming gain follows the non-central $\chi^2$-distribution (see \textit{Theorem~\ref{thm06}}), and the Cher-LB can be adopted to conduct power adaptation for a given reliability constraint.
Numerical results show that the Cher-LB significantly improves the predicted beamforming gain compared to using maximal ratio combining (MRC) alone{, since the proposed Cher-LB enables the utilization of transmit-antenna diversities.} 
While guaranteeing the outage requirement, the Cher-LB also demonstrates reasonable average power consumption {compared to the case of perfect channel knowledge}.


\section{The Chernoff Lower Bound and Analysis}\label{secII}
\subsection{Cher-LB for Non-Central $\chi^2$-Distribution}\label{secIIa}
The mathematical foundation of our Cher-LB lies in the following well-known result for the cumulative distribution function (CDF):
\begin{lem}[Chernoff upper-bound from \cite{Boucheron13}]\label{lem01}
The generic Chernoff upper-bound of the CDF for a RV (i.e., $\beta$) is given by
\begin{equation}\label{eqn17}
\int_{0}^{\beta_{\mathrm{T}}}f_\beta(x)dx\leq\inf_{\nu>0}s(\nu,\beta_{\mathrm{T}}),
\end{equation}
where 
\begin{equation}\label{eqn18}
s(\nu,\beta_{\mathrm{T}})\triangleq\exp(\nu\beta_{\mathrm{T}})\mathbb{E}(\exp(-\nu\beta)).
\end{equation}
Given the definition of $\beta$ in \eqref{eqn01}, $s(\nu,\beta_{\mathrm{T}})$ becomes the product of several individual terms:
\begin{equation}\label{eqn19}
s(\nu,\beta_{\mathrm{T}})=\exp(\nu\beta_{\mathrm{T}})\prod_{k=0}^{K-1}\mathbb{E}(\exp(-\nu\alpha_k^2)).
\end{equation}
\end{lem}

Hence, there exists a lower bound $\beta^\perp_\text{Cher}<\beta_{\mathrm{T}}$ such that
\begin{equation}\label{eqn20}
\epsilon=\inf_{\nu>0}s(\nu,\beta^\perp_\text{Cher}),
\end{equation} 
where $\beta^\perp_\text{Cher}$ is the Cher-LB of interest. 
For the notation simplificity, we drop the subscript $_\text{Cher}$ in the theoretical analysis of Sec.~\ref{secIIa} and Sec.~\ref{secIIIa}.

{Prior to touching the real problem \eqref{eqn20}, it is important to acquire $\mathbb{E}(\exp(-\nu\alpha_k^2)$ in \eqref{eqn19}. 
Fortunately, this is a reversed version of the moment generating function. 
After some tedious mathematical work, its closed-form is given by \cite{AP91}}
\begin{equation}\label{eqn21}
\mathbb{E}(\exp(-\nu\alpha_k^2))=\frac{1}{\sqrt{1+2\sigma^2\nu}}\exp\left(-\frac{\mu_k^2\nu}{1+2\sigma^2\nu}\right).
\end{equation}

With \eqref{eqn21}, $s(\nu,\beta^\perp)$ can now be written into
\begin{equation}\label{eqn22}
s(\nu,\beta^\perp)=\frac{1}{(1+2\sigma^2\nu)^{\frac{K}{2}}}\exp\left(\nu\beta^\perp-\frac{\nu\mathcal{M}^2}{1+2\sigma^2\nu}\right).
\end{equation}
It is worth noting that \eqref{eqn20} is still a transcendental problem for $\beta^\perp$ with \eqref{eqn22}.
Nevertheless, applying \eqref{eqn22} into \eqref{eqn20} with convexity analysis leads to the following conclusion: 
\begin{thm}\label{thm05}
Given $\epsilon\in(0,1)$, the Cher-LB, i.e., $\beta^\perp$, is a monotonically increasing function of $\epsilon$ and falls into the interval $(0,\mathcal{M}^2+K\sigma^2)$. 
Hence, $\beta^\perp$ can be obtained through line searching.
\end{thm}

\begin{IEEEproof}
Given $\nu>0$, \eqref{eqn22} show that $s(\nu, \beta^\perp)$ increases monotonically with $\beta^\perp$. 
Hence, the minimum of $s(\nu,\beta^\perp)$ increases with $\beta^\perp$ and vice versa.

To find the interval where $\beta^\perp$ falls into, we are interested in the partial derivative of $s(\nu,\beta^\perp)$ to $\nu$, which yields
{\begin{IEEEeqnarray}{rl}
\frac{\partial s(\nu,\beta^\perp)}{\partial \nu}=&s(\nu,\beta^\perp)\Bigl(\beta^\perp-\Bigl(\frac{\mathcal{M}^2}{(1+2\sigma^2\nu)^2}+\frac{K\sigma^2}{1+2\sigma^2\nu}\Bigl)\Bigr)\IEEEnonumber\\
\mathop{>}^\text{(c)}&s(\nu,\beta^\perp)(\beta^\perp-(\mathcal{M}^2+K\sigma^2)).\label{eq24}
\end{IEEEeqnarray}}
The inequality (c) holds due to $(1+2\sigma^2\nu)>1$.
Then, we consider two cases of interval for $\beta^\perp$:

\textit{Case 1}: $\beta^\perp\in[\mathcal{M}^2+K\sigma^2, \infty)$. 
It is  immediately followed by: $\partial s(\nu,\beta^\perp)/\partial \nu>0$, and thus $s(\nu,\beta^\perp)$ is a monotonically increasing function of $\nu$. Then, the following result holds
\begin{equation}\label{eq25}
\inf_{\nu>0}s(\nu,\beta^\perp)=\lim_{\nu\to0}s(\nu,\beta^\perp)=1.
\end{equation}
Applying \eqref{eq25} into \eqref{eqn20} gives
\begin{equation}
\epsilon=\inf_{\nu>0}s(\nu,\beta^\perp)\geq 1.
\end{equation}
This result is not in line with the fact $\epsilon\in(0, 1)$, and thus \textit{Case 1} is not valid. 

\textit{Case 2}: $\beta^\perp\in(0, \mathcal{M}^2+K\sigma^2)$.
Let $\partial s(\nu,\beta^\perp)/\partial \nu=0$ and we obtain
\begin{equation}\label{eq26}
\beta^\perp-\left(\frac{\mathcal{M}^2}{(1+2\sigma^2\nu)^2}+\frac{K\sigma^2}{1+2\sigma^2\nu}\right)=0.
\end{equation}
\eqref{eq26} is a quadratic polynomial equation. 
Since $\nu>0$, \eqref{eq26} has only one solution as given by
\begin{equation}\label{eqn25}
\nu^\star=\frac{K\sigma^2+\sqrt{K^2\sigma^4+4\beta^\perp\mathcal{M}^2}}{4\sigma^2\beta^\perp}-\frac{1}{2\sigma^2}.
\end{equation}

\eqref{eq26} also indicates
\begin{equation}\label{eq261}
\lim_{\nu^\star\rightarrow \infty}\beta^\perp=0.
\end{equation}
Moreover, we have 
\begin{equation}\label{eq27}
\lim\limits_{ \nu^\star\to\infty}\beta^\perp \nu^\star=\lim\limits_{\ \nu^\star\to\infty}\left(\frac{\mathcal{M}^2\nu}{(1+2\sigma^2\nu)^2}+\frac{K\sigma^2\nu}{1+2\sigma^2\nu}\right)=\frac{K}{2}.
\end{equation}
By substituting \eqref{eq27} to \eqref{eqn22}, we obtain
\begin{equation}\label{eq271}
\lim_{\beta^\perp\rightarrow 0}s(\beta^\perp,\nu^\star)=0,
\end{equation}
or equivalently we have $\epsilon\rightarrow 0$ when $\beta^\perp\rightarrow 0$.
Moreover, when $\beta^\perp\rightarrow(\mathcal{M}^2+K\sigma^2)$, we have $\nu^\star\to0$. 
In this case, \eqref{eqn22} gives
\begin{equation}\label{eq272}
\lim_{\nu^\star\rightarrow 0}s(\beta^\perp,\nu^\star)=1.
\end{equation}
In conclusion, as $\epsilon$ increases from $0$ to $1$, $\beta^\perp$ increases from $0$ to $(\mathcal{M}^2+K\sigma^2)$. {\em Theorem \ref{thm05}} is therefore proved. 
\end{IEEEproof}

The line searching algorithm is provided in \textbf{Algorithm~\ref{agthm2}}, which aims to obtain an estimate of $\beta^\perp$ (i.e., $\hat{\beta}^\perp$).
Specifically, $\text{sgn}$ stands for the sign-function.
The complexity to find $\beta^\perp$ is $\mathcal{O}(\lceil\log_{2}(\frac{(\mathcal{M}^2+K\sigma^2)}{\delta_{\beta}})\rceil$.
For instance, when $\frac{(\mathcal{M}^2+K\sigma^2)}{\delta_{\beta}}$ is nearly $128$, the complexity is $\mathcal{O}(7)$.
Compared to the approximations, the Abdel-Aty's closer approximation has the highest complexity, involving three sixth-order polynomials (i.e., the complexity is roughly $\mathcal{O}(3\log_{2}(6))=\mathcal{O}(7.6)$); while the Sankaran's $z1$ approximation has the lowest complexity, involving only two quadratic polynomials (i.e., the complexity is $\mathcal{O}(2\log_{2}(2))=\mathcal{O}(2)$). 
Hence, the complexity of \textbf{Algorithm~\ref{agthm2}} is comparable to the aforementioned approximation approaches, and is suitable for real-time signal processing. 

Notably, when $\mathcal{M}^2=0$, \eqref{eqn22} and \eqref{eqn25} reduce to
\begin{equation}\label{eqn27a}
s(\nu,\beta^\perp)=\frac{\exp(\nu\beta^\perp)}{(1+2\sigma^2\nu)^{\frac{K}{2}}};~\nu^\star=\frac{K}{2\beta^\perp}-\frac{1}{2\sigma^2}.
\end{equation}
After some tidy-up work, \eqref{eqn27a} is equivalent to the discussion in \cite{9120745} (i.e., the central $\chi^2$ distribution case).

\subsection{Numerical Results of Cher-LB}\label{secIId}
In this subsection, numerical results are employed to demonstrate the advantages of the Cher-LB {in comparison to the Poly-LB and the approximations.
The First and Closer approximations in \cite{AbdelAty1954} approximate $\sqrt[3]{\beta}$ after scaling to be Gaussian, while the $z1$ and $z2$ approximations in \cite{Sankaran1963} use $\sqrt{\beta}$.
}

\begin{figure}
\centering
\includegraphics[scale=0.43]{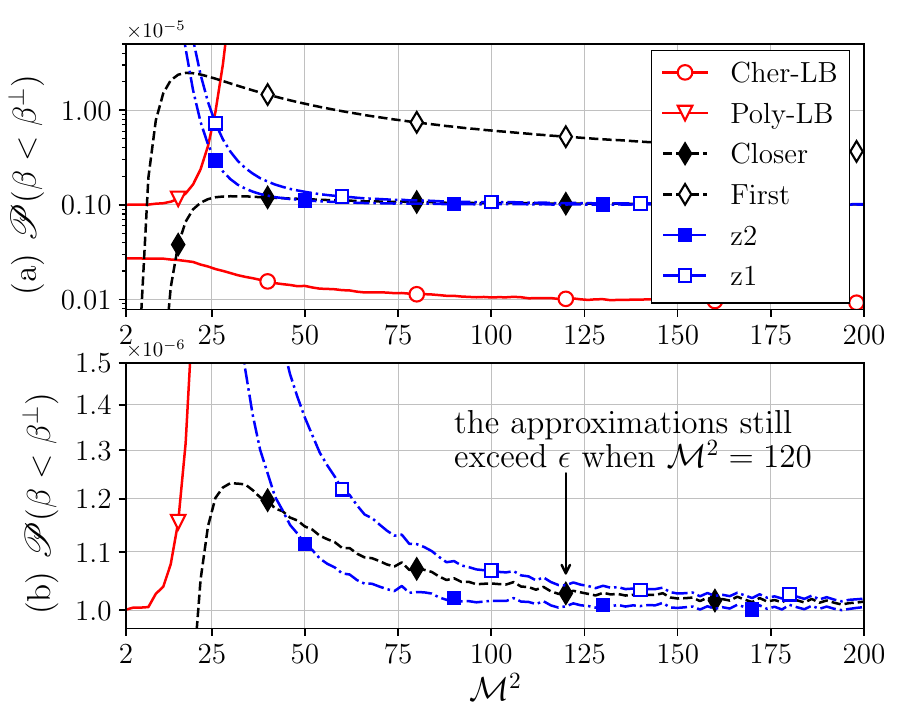}
\caption{Outage probability of lower-bounds and approximations as a function of $\mathcal{M}^2$ ($\sigma^2=1$, $K=4$ and $\epsilon=10^{-6}$).}
\vspace{-1.0em}
\label{fig3NonCentral}
\end{figure}

{
	\begin{algorithm}[t]
		\caption{\footnotesize Line Searching for $\hat{\beta}^\perp$ (Non-Central $\chi^2$)}
		\begin{algorithmic}[1]\label{agthm2}\footnotesize
			\renewcommand{\algorithmicrequire}{\textbf{Input:}}
			\renewcommand{\algorithmicensure}{\textbf{Output:}}
			\REQUIRE Outage-probability constraint $\epsilon$, $\mathcal{M}^2$, $\sigma^2$, tolerance $\delta_{\beta}$;
			\ENSURE  Cher-LB $\hat{\beta}^\perp$;
			\STATE \textbf{Initialization}: $\beta_\text{low}=0$, $\beta_\text{up}=\mathcal{M}^2+K\sigma^2$; 
			\WHILE{$\beta_\text{up}-\beta_\text{low}>\delta_{\beta}$}
			\STATE $\beta_\text{mid}\leftarrow(\beta_\text{low}+\beta_\text{up})/2$;
			\STATE Calculate $\nu^\star$ based on $\beta_\text{mid}$ and \eqref{eqn25};
			\STATE $\beta_\mathrm{s}\leftarrow(\text{sgn}(s(\nu^\star,\beta_\text{mid})-\epsilon)+1)/2$;
			\STATE $\beta_\text{up}\leftarrow(1-\beta_\mathrm{s})\beta_\text{mid}+\beta_\mathrm{s}\beta_\text{up}$; $\beta_\text{low}\leftarrow\beta_\mathrm{s}\beta_\text{mid}+(1-\beta_\mathrm{s})\beta_\text{low}$;
			\ENDWHILE
			\STATE $\hat{\beta}^\perp\leftarrow\beta_\text{low}$;
			\RETURN $\hat{\beta}^\perp$. 
		\end{algorithmic} 
	
	\end{algorithm}
	
}



The first thing of interest is the effectiveness of the Cher-LB, as shown in Fig.~\ref{fig3NonCentral}.
The variance of each Gaussian RV is normalized to $\sigma^2=1$. 
Moreover, the DoF is set to $K=4$ and outage requirement is $\epsilon=10^{-6}$ (related to the URLLC use cases addressed in Sec.~\ref{secIIId}). 
The first thing to be noticed in Fig.~\ref{fig3NonCentral}(a) is that the Cher-LB maintains an outage probability below $\epsilon$ for all values of $\mathcal{M}^2$, distinguishing itself as the only method to do so.
With the increase of $\mathcal{M}^2$, its outage probability remains around $10^{-7}$, which is not sensitive to the variation of $\mathcal{M}^2$.
On the other hand, the outage probability of the Poly-LB quickly exceeds $\epsilon$ with the increase of $\mathcal{M}^2$.
Upon examining the approximations, it is observed that they all show significant inaccuracy when $\mathcal{M}^2$ is small.
Moreover, all approximations exceeds the outage requirement $\epsilon$ when $\mathcal{M}^2>20$.
This means the approximations cannot enable URLLC services alone, where guaranteeing the expected outage probability is of utmost importance.
Fig.~\ref{fig3NonCentral}(b) provides a closer look at the approximations, and shows that the approximations still exceeds $\epsilon$ by $2\%\sim5\%$ when $\mathcal{M}^2=120$.
Such increase of error is significant for mission-critical services.
In Sec.~\ref{secIIId}, it will be demonstrated that there is a high probability of $\mathcal{M}^2<120$ in MIMO systems.

\begin{figure}[t!]
\centering
\includegraphics[scale=0.43]{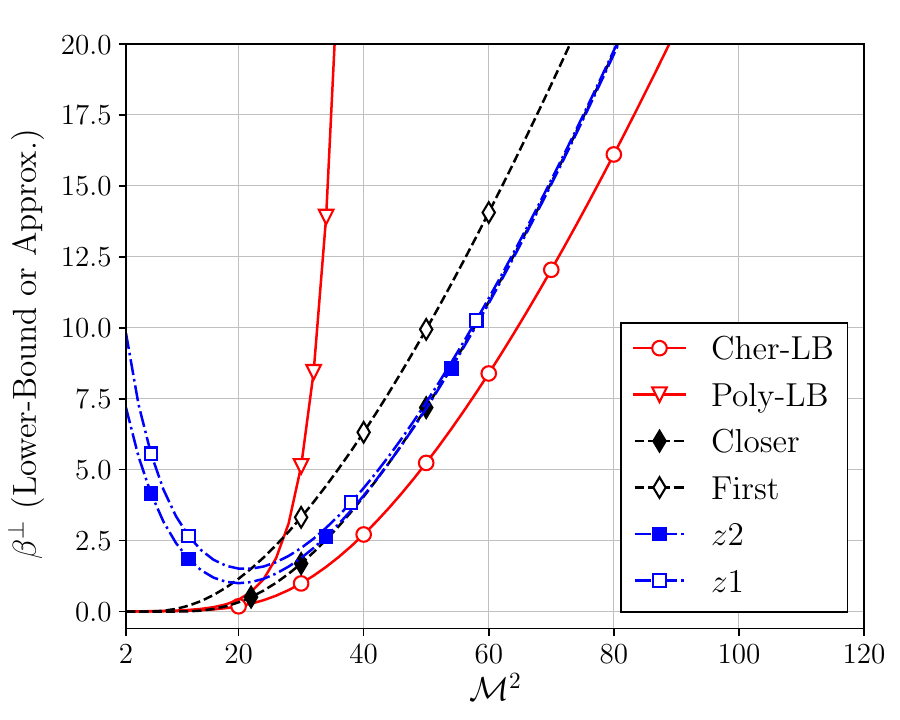}
\caption{Lower-bounds and approximations of $\beta_{\mathrm{T}}$ as a function of $\mathcal{M}^2$ ($\sigma^2=1$, $K=4$ and $\epsilon=10^{-6}$).}
\vspace{-2.1em}
\label{fig1NonCentral}
\end{figure}

After examining the effectiveness, we further demonstrate the closeness of the Cher-LB to the approximations, which gradually converge to $\beta_{\mathrm{T}}$ as $\mathcal{M}^2$ increases, as shown in Fig.~\ref{fig1NonCentral}.
It is shown that when $\mathcal{M}^2>40$, the Cher-LB has been reasonably close to the convergence of the $z2$, $z1$ and closer approximations (i.e., around $1.25$ less than the approximations).
Moreover, as $\mathcal{M}^2$ increases, the difference between the Cher-LB and the approximations only exhibits slight increases.
This means the Cher-LB becomes closer to $\beta_{\mathrm{T}}$ in terms of percentage as $\mathcal{M}^2$ increases.
This means the Cher-LB achieves reasonable tightness while always fulfilling the outage requirement.


\section{Application of Cher-LB for MIMO-URLLC in The First-Order Markov Channel}\label{secIII}
\subsection{System Model and MIMO-URLLC Problem}\label{secIIIa}
Consider a narrowband MIMO-URLLC system, where an access point (AP) with $M$ transmit-antennas aims to communicate to users with an ultra-low outage probability (i.e., $\epsilon$).
Concerning the MIMO channel to be slowly or moderately time-varying, it can be mathematically described by the first-order Markov model (see \cite{1593619})
\begin{equation}\label{eqn35}
\mathbf{H}_{t+\tau}=\mathcal{J}_0(2\pi f_\text{d}\tau)\mathbf{H}_{t}+\boldsymbol{\Omega}_{t+\tau},
\end{equation}
where $\mathbf{H}\in\mathbb{C}^{N\times M}$ stands for the MIMO channel matrix \footnote{Here $N$ is the number of receive antennas which can be either co-located at a single user or distributed at different users. Later on, we will show that the DoF of the beamforming gain is $K=2N$ as in Section \ref{secIIIb}.}, $t$ for the time index,  $\tau$ for the lag, $\mathcal{J}_0(\cdot)$ for the zero-order Bessel function of the first kind, $f_\text{d}$ for the maximum Doppler shift, and $\boldsymbol{\Omega}_{t+\tau}\in\mathbb{C}^{N\times M}$ for i.i.d. complex Gaussian matrix. 
Each element of $\boldsymbol{\Omega}_{t+\tau}$ follows the distribution $\mathcal{CN}(0,\sigma_\omega^2)$ with $\sigma_\omega^2=1-\mathcal{J}^2_0(2\pi f_\text{d}\tau)$. 
Since the term $(2\pi f_\text{d})$ keeps constant throughout the rest of this section, it is abbreviated in the Bessel function for the sake of notation simplicity, i.e., $\mathcal{J}_0(\tau)\triangleq\mathcal{J}_0(2\pi f_\text{d}\tau)$.

In this context, the AP (i.e., the transmitter) employs multi-antenna precoding/beamforming for the spatial multiplexing and/or spatial diversity.
To this end, the transmitter requires the CSI-T and captures it at the time slot $t$, i.e., the transmitter knows $\mathbf{H}_t$, 
and based on which the precoding/beamforming matrix $\mathbf{W}\in\mathbb{C}^{M\times {L}}$ is formed (${L}\leq N$ stands for the number of data streams).
At the time of transmission (i.e., at the time slot $t+\tau$), \eqref{eqn35} shows that the actual CSI becomes $\mathbf{H}_{t+\tau}$, where $\mathbf{\Omega}_{t+\tau}$ is the the CSI-T uncertainty.
Therefore, the received signal at receiver(s), denoted by $\mathbf{y}_{t+\tau}$, is
\begin{IEEEeqnarray}{rl}
\mathbf{y}_{t+\tau}&=\mathbf{H}_{t+\tau}\mathbf{W}\mathbf{s}+\mathbf{v}\label{eqn36}\\
&=\mathcal{J}_0(\tau)\mathbf{H}_t\mathbf{W}\mathbf{s}+\mathbf{\Omega}_{t+\tau}\mathbf{W}\mathbf{s}+\mathbf{v},\label{eqn37}
\end{IEEEeqnarray} 
where $\mathbf{s}\in\mathbb{C}^{{L}\times1}$ stands for the transmitted symbol-block with $\mathbb{E}(\mathbf{s})=\mathbf{0}$ and $\mathbb{E}(\mathbf{s}\mathbf{s}^H)=E_s\mathbf{I}_{L}$ ($\mathbf{I}_{L}$: the identity matrix with the size ${L}$; $E_s$: the symbol energy), 
and $\mathbf{v}\in\mathbb{C}^{N\times1}$ for the additive white-Gaussian noise (AWGN) with $\mathbf{v}\sim\mathcal{CN}(\mathbf{0}, \sigma_v^2\mathbf{I}_N)$.

Consider the use of interference-rejection precoding (see \cite{1468466}), i.e., $\mathbf{H}_t\mathbf{W}=\mathbf{I}_{N}$.
To simplify our presentation, here assumes ${L}=N$ and rewrite \eqref{eqn37} into
\begin{equation}\label{eqn38}
\mathbf{y}_{t+\tau}=\mathcal{J}_0(\tau)\mathbf{s}+\mathbf{\Omega}_{t+\tau}\mathbf{W}\mathbf{s}+\mathbf{v}.
\end{equation} 
Denote $\boldsymbol{\omega}_n^T$ to be the $n^{th}$ row of $\mathbf{\Omega}_{t+\tau}$ and $\mathbf{w}_n$ to be the $n^{th}$ column of $\mathbf{W}$; the superscript $[\cdot]^T$ stands for the matrix/vector transpose.
The signal-to-interference and noise ratio (SINR) for the $n^{th}$ element of $\mathbf{y}_{t+\tau}$ is computed by
\begin{IEEEeqnarray}{ll}\label{eqn39}
\textsc{sinr}_n&=\frac{|\mathcal{J}_0(\tau)+\boldsymbol{\omega}_n^T\mathbf{w}_n|^2E_s}{\sigma_v^2+{\Upsilon_n}}~_{{l},n\in\{0,...,N-1\}}\\
&\leq \frac{\mathcal{J}^2_0(\tau)E_s{+2\mathcal{J}_0(\tau)|\boldsymbol{\omega}_n^T\mathbf{w}_n|E_s}+|\boldsymbol{\omega}_n^T\mathbf{w}_n|^2E_s}{\sigma_v^2+{\Upsilon_n}},~~~~\label{eqn40}
\end{IEEEeqnarray}
where ${\Upsilon_n}=\sum_{l\neq n}|\boldsymbol{\omega}_n^T\mathbf{w}_{l}|^2E_s$ is the inter-stream interference.
Hence, current interference-rejection precoding techniques cannot eliminate the interference caused by the CSI-T uncertainty $\mathbf{\Omega}_{t+\tau}$.
The interference ${\Upsilon_n}$ is detrimental to URLLC users mainly in the sense that: 
{\em 1)} ${\Upsilon_n}$ grows $(N-1)$ times faster than the potentially contributive term $|\boldsymbol{\omega}_n^T\mathbf{w}_n|^2E_s$, and thus SINR monotonically decreases with respect to $N$; 
{\em 2)} when $\tau$ and/or $f_\text{d}$ are/is considerably large, SINR would be dominated by the interference. 
In this case, SINR reads as
\begin{equation}\label{eqn41}
\textsc{sinr}_n\approx\frac{|\mathcal{J}_0(\tau)+\boldsymbol{\omega}_n^T\mathbf{w}_n|^2}{\sum_{{l}\neq n}|\boldsymbol{\omega}_n^T\mathbf{w}_{l}|^2}.
\end{equation}
Since the SINR term \eqref{eqn41} is independent of $E_s$, it is not possible to improve the communication reliability by means of adapting the transmit-energy $E_s$. 
Then, we will have to reduce the transmission rate (i.e., increase transmission redundancy) in order to improve the reliability. 
This is however by means of trading off the spectral efficiency and more critically the latency, which is not a suitable approach for URLLC.

One might propose to use massive-MIMO because  $\lim\limits_{M\to\infty}\boldsymbol{\omega}_n^T\mathbf{w}_{l}=0, \forall {l},n,$
with which we have the so-called channel hardening effect
\begin{equation}\label{eqn43}
\lim_{M\to\infty}\textsc{sinr}_n=\frac{\mathcal{J}_0^2(\tau)E_s}{\sigma_v^2}.
\end{equation}
We argue that \eqref{eqn43} hold not only for the interference-rejection precoding but also for the very simple matched-filter (MF) beamforming approach, i.e., $\mathbf{W}=\mathbf{H}_t^H$ ($[\cdot]^H$ denotes the matrix/vector Hermitian transpose).
Therefore, the real problem lies in the scope of not-too-large MIMO, where the interference term in \eqref{eqn39}-\eqref{eqn40} is an issue.  
In practice, interference-avoidance strategy is employed, i.e., users (or streams) are allocated on different frequencies (subchannels). 
By this means, our system of interest becomes single-stream point-to-point MIMO (${L}=1$), where the inter-stream interference is avoided. 

\subsection{Prediction of Beamforming Gain in MIMO-URLLC}\label{secIIIb}
Given the system setup of ${L}=1$ and $N>{L}$, \eqref{eqn36} becomes 
\begin{equation}\label{eqn44}
\mathbf{y}_{t+\tau}=s\mathbf{H}_{t+\tau}\mathbf{W}\mathbf{1}_N+\mathbf{v},
\end{equation}
where $\mathbf{1}_N$ denotes an $(N)\times (1)$ $1$-vector. 
It is reasonable to assume that the receiver knows $\mathbf{H}_{t+\tau}$ and $\mathbf{W}$. 
Therefore, a receiver beamforming-vector $\mathbf{u}=\mathbf{H}_{t+\tau}\mathbf{W}\mathbf{1}_N$ can be employed to enable the maximum-ratio combine (MRC) as
\begin{IEEEeqnarray}{ll}\label{eqn45}
y&=\mathbf{u}^H\mathbf{y}_{t+\tau}\\
&=\|\mathbf{H}_{t+\tau}\mathbf{W}\mathbf{1}_N\|^2s+\tilde{v},\label{eqn46}
\end{IEEEeqnarray}
where $\tilde{v}$ is the corresponding AWGN after the receiver beamforming.  
The signal-to-noise ratio (SNR) is computed by
\begin{equation}\label{eqn47}
\textsc{snr}=\frac{\|\mathbf{H}_{t+\tau}\mathbf{W}\mathbf{1}_N\|^2E_s}{\sigma_v^2}.
\end{equation}
Hence, the beamforming gain of this system is given by
\begin{IEEEeqnarray}{ll}\label{eqn48}
\beta&=\|\mathbf{H}_{t+\tau}\mathbf{W}\mathbf{1}_N\|^2\\
&=\|\mathcal{J}_0(\tau)\mathbf{H}_t\mathbf{W}\mathbf{1}_N+\mathbf{\Omega}_{t+\tau}\mathbf{W}\mathbf{1}_N\|^2.\label{eqn49}
\end{IEEEeqnarray} 

\begin{thm}\label{thm06}

Let $\boldsymbol{\alpha}=\mathbf{H}_{t+\tau}\mathbf{W}\mathbf{1}_N$.
Then, $\beta$ is the sum of squares of $2N$ real-Gaussian RVs: $\beta=\sum_{n=0}^{N-1}(\Re(\alpha_n)^2+\Im(\alpha_n)^2)$.
$\Re(\alpha_n)$ and $\Im(\alpha_n)$, $_{n=0,1,\cdots N-1}$, have identical variance:
\begin{equation}\label{eqn50}
\sigma^2=\sigma_{\omega}^2\|\mathbf{W}\mathbf{1}_N\|^2/2.
\end{equation}
The mean of $\alpha_n$ is given by
\begin{equation}\label{eqn51}
\mu_n=\mathcal{J}_0(\tau)\mathbf{h}^T_{t,n}\mathbf{W}\mathbf{1}_N,
\end{equation}
where $\mathbf{h}^T_{t,n}$ is the $n^{th}$ row vector of $\mathbf{H}_{t}$.
Hence, the mean of $\Re(\alpha_n)$ and $\Im(\alpha_n)$ is given by $\Re(\mu_n)$ and $\Im(\mu_n)$, respectively.

According to the definition in \eqref{eqn01}, $\beta$ obeys the non-central $\chi^2$-distribution with $K=2N$ and 
\begin{equation}\label{eqn52}
\mathcal{M}^2=\mathcal{J}_0^2(\tau)\sum_{n=0}^{N-1}\|\mathbf{h}^T_{t,n}\mathbf{W}\mathbf{1}_N\|^2.
\end{equation}

\end{thm}
\begin{IEEEproof}
The proof is rather straightforward and thus omitted for the sake of spacing limit. 
\end{IEEEproof}

{\em Theorem \ref{thm06}} shows that $\beta$ is a non-central $\chi^2$-distributed RV. 
Given the reliability constraint $\epsilon$, we can use {\em Theorem~\ref{thm05}} and {\bf Algorithm \ref{agthm2}} to obtain the Cher-LB, $\beta^\perp$.
This helps the transmitter to know 
\begin{equation}\label{eqn53}
\mathscr{P}\Bigl(\textsc{snr}\geq\frac{\beta^\perp E_s}{\sigma_v^2}\Bigl)\geq1-\epsilon.
\end{equation}
and scales $E_s$ accordingly to guarantee the reliability.

Compared to using MRC alone, \textit{Theorem~\ref{thm06}} enables the exploitation of transmit-antenna diversities.
This motivates us to study the behavior in the case of massive-MIMO (i.e., $M\to\infty$).
Given that $\mathbf{W}$ is uncorrelated with $\mathbf{\Omega}_{t+\tau}$, we immediately have
\begin{equation}\label{eqn55}
\lim_{M\to\infty}\beta=\mathcal{J}_0^2(\tau)\|\mathbf{H}_t\mathbf{W}\mathbf{1}_N\|^2\mathop{\leq}^\text{(a)}\mathcal{J}_0^2(\tau),
\end{equation}
where the equality on {(a)} holds when $\mathbf{W}$ is designed to fulfill $\|\mathbf{H}_t\mathbf{W}\mathbf{1}_N\|^2=1$.
This result is in line with \eqref{eqn43}.

Unsurprisingly, in the presence of CSI-T uncertainty, we cannot improve the beamforming gain from the receiver side. 
However, by taking advantage of transmit-antenna diversities, the predicted beamforming gain is anticipated to gain significant improvement compared to using MRC alone.


\subsection{Numerical Results and Discussion}\label{secIIId}
In this subsection, numerical results are employed to demonstrate the performance of the Cher-LB in MIMO systems under first-order Markov channel and elaborate our theoretical results. As discsussed in Section \ref{secIIIa}, we consider single-stream point-to-point MIMO, where the receiver (i.e., user) only has a small number of antennas. $\mathbf{H}_t$ is assumed to be i.i.d. Rayleigh \cite{8660712}, and MF beamforming (i.e., $\mathbf{W}=\mathbf{H}_t^H$) is adopted in appreciation of its low computational complexity. Since the beamforming gain is normalized, $\delta_{\beta}$ of {\textbf{Algorithm~\ref{agthm2}}} is reduced to {$10^{-10}$}. The central carrier frequency is assumed to be $3.5$ GHz (see \cite{Euro5GObservatory}). Considering the stringent latency requirement of URLLC, the time lag $\tau$ is set to be $0.5$ ms. The user velocity is assumed to be $20$ m/s (i.e., $72$ km/h), which is a pessimistic consideration of vehicular velocity in urban areas.
Monte-Carlo trials are performed to study the properties of interest. 

\begin{figure}[t]
\centering
\includegraphics[scale=0.43]{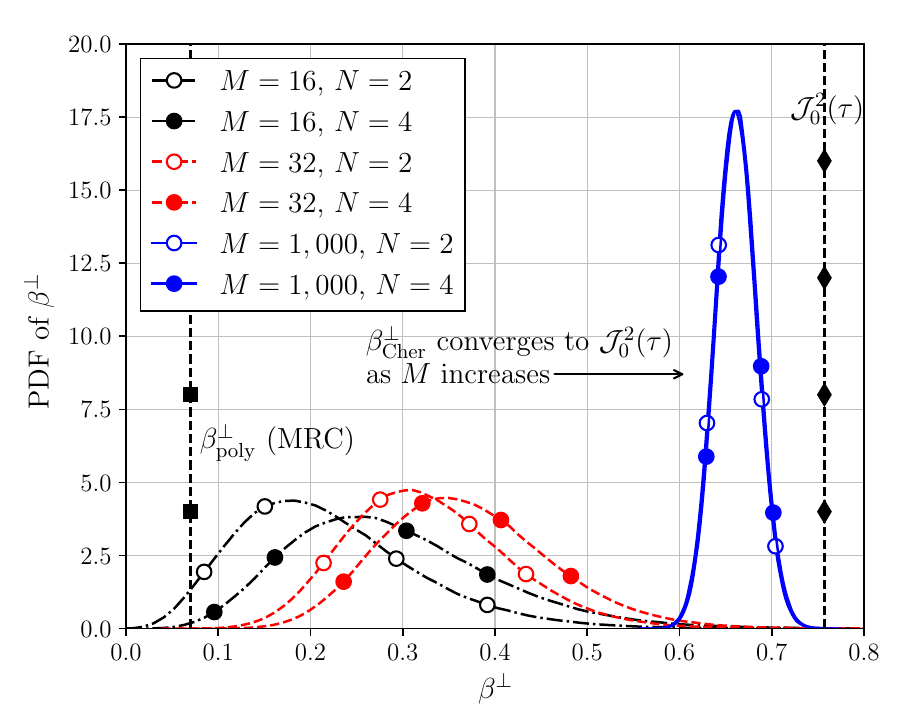}
\caption{The PDF of $\beta^\perp_\text{Cher}$ as $M$ increases from $16$ to $32$ and $1,000$ when using MF beamforming, $\epsilon=10^{-6}$, $N=2$ or $4$.}
\vspace{-2.1em}
\label{fig7}
\end{figure}

The first thing of interest is to compare the predicted beamforming gain to using MRC alone, as shown in Fig.~\ref{fig7}
We consider the case where $N=2$ (marked by hollow circles) or $4$ (marked by solid circles). 
For MRC, the Poly-LB of $N=4$ (i.e., $K=8$) is presented in appreciation of its good tightness.
It is observed that using beamforming gain prediction for MF beamforming significantly outperforms using MRC alone.
When $M=16$, $\beta^\perp_{\text{Cher}}$ achieves around $0.17$, compared to $\beta^\perp_{\text{poly}}=0.07$.
Moreover, as $M$ increases to $32$, $\beta^\perp_{\text{Cher}}$ increases to around $0.3$.
This is because a larger $M$ brings more beamforming gain (i.e., higher ratio of $(\mathcal{M}^2)/(\sigma^2)$).
As $M$ increases to $1,000$, $\beta^\perp_\text{Cher}$ approaches the limit of $\mathcal{J}_0^2(\tau)$. 
This coincides with the discussions in Section \ref{secIIIb}.  
In addition, increasing $N$ also improves $\beta^\perp_{\text{Cher}}$ when $M$ is not too large.
This is because $\mathbf{H}_t$ becomes quasi-orthogonal with the increase of $M$ (e.g., $M=1,000$), and increasing $N$ makes little difference in this case.
Nevertheless, when $M$ is $16$ or $32$, the improvement of increasing $N$ is significant ($50\%$ or $20\%$, respectively).

\begin{figure}[t]
	\centering
	\includegraphics[scale=0.43]{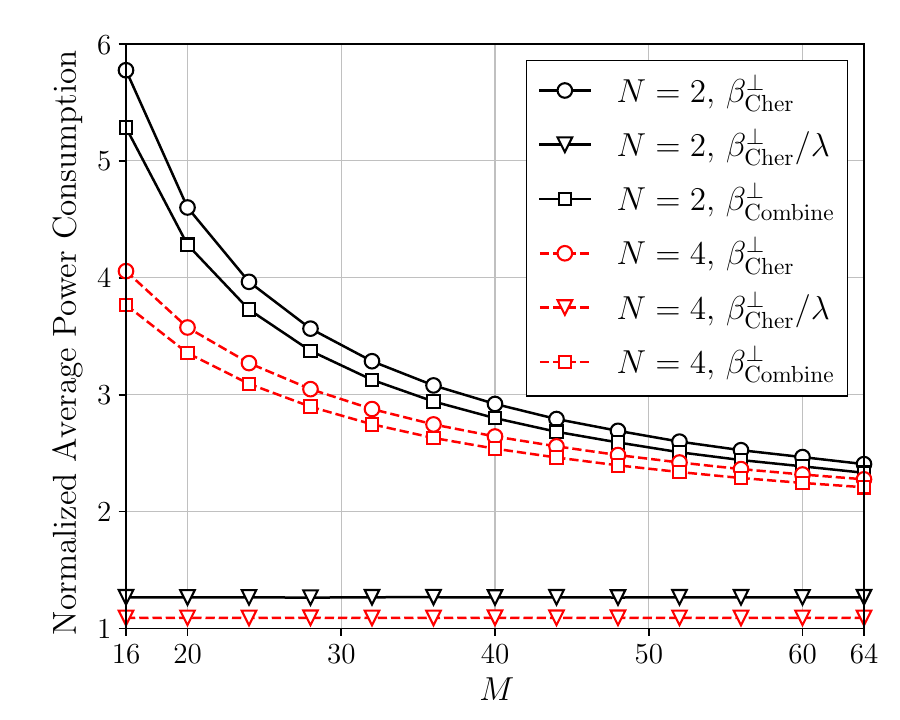}
	\caption{The normalized average power consumption for power adaptation when $\epsilon=10^{-6}$, $M$ increases from $16$ to $64$.}
	\vspace{-2.1em}
	\label{fig8}
\end{figure}

Secondly, we are interested in the average power consumption of the power adaptation based on the predicted beamforming gain, as shown in Fig.~\ref{fig8}.
With a fixed SNR target, $E_\text{s}$ linearly scales to $(1)/(\beta^\perp_\text{Cher})$.
Hence, we use $\mathbb{E}((1)/(\beta^\perp_\text{Cher}))$ as the normalized average power consumption. 
In addition to using the Cher-LB alone, we consider a combination of the Cher-LB and the $z2$ approximation, where $z2$ approximation is deemed to be close enough to $\beta_{\mathrm{T}}$ when $(\mathcal{M}^2)/(\sigma^2)>120$ based on Fig.~\ref{fig3NonCentral}.
This approach helps to demonstrate the improvement on $\beta^\perp_\mathrm{Cher}$ if $\beta_{\mathrm{T}}$ is available for limited occasions.
These approaches are compared to an optimistic case where the transmitter exactly knows $\beta$, i.e., the normalized average power consumption is $\mathbb{E}((1)/(\beta))$.

Compared to $\mathbb{E}((1)/(\beta))$, which overall stays around $1.25$, $\mathbb{E}((1)/(\beta^\perp_{\text{Cher}}))$ decreases from around $5.5$ to $2.2$ as $M$ increases from $16$ to $64$.
{This improvement is because of the increase of transmit-antenna diversities.
Moreover, this means a $75\%$ increase of power consumption compared to the case of perfect channel knowledge.
Considering the extreme reliability requirement, this is a reasonable performance.}
Moreover, the combination approach only shows around $0.2$ improvement compared to using the Cher-LB alone.
This is reasonable, since when $(\mathcal{M}^2)/(\sigma^2)>120$, the Cher-LB is already close to the approximations.
Therefore, using the Cher-LB alone can achieve reasonable average power consumption.

\section{Conclusion}\label{secIV}
In this paper, we have investigated lower bounds of the outage threshold when RV obeys the non-central $\chi^2$ distribution.
It has been shown, through both theoretical and numerical analysis, that Cher-LB is the most {effective} lower bound amongst all studied candidates. 
The closed-form of Cher-LB was found hard to derive. 
Nevertheless, it could be obtained through line searching over a specific domain of the function. 
As far as MIMO-URLLC is concerned for the application, the Cher-LB has been employed to form pessimistic prediction of the MIMO transmitter beamforming-gain when the CSI-T obeys the first-order Markov process. 
It has been shown that the pessimistic prediction is made sufficiently accurate for the guaranteed reliability as well as the transmit-energy efficiency.

{
\section*{Acknowledgement}
This work is partially funded by the 6G Innovation Centre.
}

\balance

\ifCLASSOPTIONcaptionsoff
\newpage
\fi

\bibliographystyle{IEEEtran}
\bibliography{URLLC,Bib_Else,Books_and_Standards,NLP_Downlink}

\end{document}